\documentclass[12pt]{article}
\usepackage{amsmath}
\usepackage{geometry}
\usepackage{algorithm,float}
\usepackage[noend]{algpseudocode}
\usepackage{graphicx}
\usepackage{caption}
\usepackage{setspace}
\usepackage[font={small,it}]{caption}
\usepackage{enumitem}
\usepackage{amssymb}
\usepackage[noend]{algpseudocode}
\usepackage{algorithm}
\usepackage{bm}
\usepackage{color}
\usepackage{epstopdf}
\usepackage{pdflscape, amsthm}
\usepackage{subcaption}
\usepackage{multirow}

\DeclareMathOperator*{\argmin}{\arg\!\min}

\newcommand{\bmX}{\bm{X}}
\newcommand{\bmx}{\bm{x}}
\newcommand{\bmW}{\bm{W}}
\newcommand{\bmw}{\bm{w}}

\newcommand{\bmu}{\bm{u}}

\title{A Simple Correction Procedure for High-Dimensional Generalized Linear Models with Measurement Error}
\author{Michael Byrd and Monnie McGee}

\begin{document}

\maketitle

\begin{abstract}
We consider high-dimensional generalized linear models when the covariates are contaminated by measurement error. Estimates from errors-in-variables regression models are well-known to be biased in traditional low-dimensional settings if the error is unincorporated. Such models  have recently become of interest when regularizing penalties are added to the estimation procedure. Unfortunately, correcting for the mismeasurements can add undue computational difficulties onto the optimization, which a new tool set for practitioners to successfully use the models. We investigate a general procedure that utilizes the recently proposed Imputation-Regularized Optimization algorithm for high-dimensional errors-in-variables models, which we implement for continuous, binary, and count response type. Crucially, our method allows for off-the-shelf linear regression methods to be employed in the presence of contaminated covariates. We apply our correction to gene microarray data, and illustrate that it results in a great reduction in the number of false positives whilst still retaining most true positives.
\end{abstract}

\section{Introduction}

Complex, high-dimensional data sets have become the norm for many fields where it is often of interest to uncover underlying structures and to estimate the effect size of a given relationship between the observed variables.  For instance, in a microarray experiment it may be of value to identify which genes are related to some quantitative outcome or if a particular gene influences presence of a disease.  Statistical regularization procedures have been essential to addressing these fundamental problems.  In particular, when the number of variables $p$ is larger than the sample size $n$, traditional methods, such as least squares regression, can no longer be used due to identifiability issues.  Hence, regularization procedures, like the Lasso \cite{tibshirani1996regression} and the Minimax Concave Penalty (MCP) \cite{zhang2010nearly}, have become necessary tools for practitioners to identify patterns in their studies for a wide variety of problems \cite{hastie2015statistical}.

For $i = 1 , \ldots , n$, consider the generalized linear model (GLM) for independent and identically distributed pairs of responses and covariates $(y_i , \bmx_i)$, such that
\begin{equation}
\mathbb{E}(y_i) = f(\bmx_i^{T} \bm\beta)
\end{equation}
for covariates, $\bm\beta \in \mathbb{R}^{p}$, and inverse-link function, $f$ \cite{mccullagh2019generalized}.  The regularized GLM aims to minimize the objective
\begin{equation}
Q(\bm\beta ; \bmX , \bm{y} , \lambda) = 
\mathcal{L}(y ; \bmx , \bm\beta) + P(\bm\beta ; \lambda)
\label{eq:glm}
\end{equation}
with respect to $\bm\beta$, where $\mathcal{L}(y ; \bmx , \bm\beta)$ is the negative log-likelihood function and  $P(\bm\beta ; \lambda)$ is a penalty function on the coefficients.  The regularization parameter $\lambda$ determines the overall level of sparsity, and is typically tuned with cross-validation \cite{friedman2001elements}.  This formulation captures most types of data, including continuous, categorical, and count.  Adaptations of the GLM have been well studied for many common penalties \cite{van2008high}, and the objective in Equation \eqref{eq:glm} has many implemented procedures for a wide variety of problems \cite{wu2008coordinate} \cite{breheny2011coordinate}.

In addition to regularized procedures' well documented empirical performance, favorable theoretical properties, such as selection consistency, have been well studied \cite{hastie2015statistical}.  These properties, however, make the assumption that the observed covariates are perfectly measured, which, in many contexts, is not a realistic assumption.  For instance, in microarray experiments there are many possible sources for random error to be incorporated naturally into the data collection process \cite{rocke2001model}.  While the Affymetrix microarray itself is manufactured under controlled conditions according to precise specifications, the genetic material prepared as the microarray sample is subject to propagation of error. RNA preparation, for example, takes at least three days, and requires up 15 steps per day. At any one of these steps, error or contaminants leading to error could be introduced \cite{perez2006preparation}.  

We consider high-dimensional variable selection and estimation for GLMs in the context of measurement error.  In particular, we address the additive measurement error setting, which is known to cause a decrease in selection and estimation quality if not corrected \cite{loh2011high} \cite{sorensen2015measurement}.  Error-in-variables (EIV) regression has been a known issue in statistics, and has been well studied in a plethera of contexts \cite{carroll2006measurement}.  The effect of mismeasured covariates in EIV regression are biased estimates of the regression coefficients and a higher Type I error rate.  An analysis that incorporates and corrects for measurement error will aim to result in consistent estimates with fewer false positives.  This correction, however, will come with decreases in power and model efficiency.

To overcome the limitation of not directly observing the variables of interest, but rather contaminated variations, we make use of the Imputation-Regularized Optimization (IRO) algorithm \cite{liang2018imputation}.  The IRO-algorithm was proposed as a technique for missing data in the high-dimensional setting, which notably gives a flexible framework with consistency guarantees for latent variables.  Recently, the IRO-algorithm was used in the context of estimating Gaussian graphical models with mismeasured observations \cite{Byrd2019}. The procedure was shown to be asymptomatically consistent; in addition it greatly reduced the number of false positives found in the selection process and reduced the overall estimation error.  Our contribution to the measurement error problem is an extension to the IRO-algorithm for common types of generalized linear models in the presence of measurement error.  

Further, the implementation is based on a simple framework for high-dimensional measurement error problems, and we implement the procedure using well established tools; thus making the procedure easy to use for practicioners. 

Our goal is to provide a simple framework for high-dimensional measurement error problems that can be implemented using well established tools and procedures.

\subsection{Literature Review} \label{IRO-GLM-LitReview}

High-dimensional EIV regression procedures have accumulated much attention due to the fact that the contaminated observations result in inconsistent estimates and poor variable selection \cite{sorensen2015measurement} \cite{nghiem2018simulation} \cite{belloni2016ell_}.  These procedures typically correct for the contamination by incorporating the assumed known or estimable measurement error variability into the optimization \cite{loh2011high} \cite{sorensen2015measurement} \cite{datta2017cocolasso}, or by some pivotal estimation without a well defined likelihood \cite{belloni2016ell_} \cite{sorensen2018covariate}.  Notably, these procedures tend to make traditionally convex penalties into non-convex formulations, requiring special care in development of optimization routines to solve them.  Moreover, even when these issues have been addressed, model tuning is known to be more difficult as standard cross-validation is not easily applied for contaminated observations \cite{datta2017cocolasso} \cite{datta2019note}.

While many of these procedures offer nice theoretical properties for symmetric, continuous responses, few have explored a more general framework for different types of response data.  We focus on two well established methods for correcting for measurement error: (1) the Corrected Lasso (CLasso) and (2) The Generalized Matrix Uncertainty Selector (GMUS).  Both CLasso and GMUS were originally established in the Gaussian error case, see \cite{loh2011high} and \cite{rosenbaum2010sparse}, but have been established in the GLM framework by \cite{sorensen2015measurement} and \cite{sorensen2018covariate}, respectively.  The CLasso attempts to account for the bias introduced with the measurement error by incorporating it into the optimization problem with two hyperparameters controlling the size of coefficients \cite{sorensen2015measurement}.  GMUS takes a slightly different approach, limiting the amount of correlation between the response and covariates by bounding the score function with a Taylor series expansion of the residual \cite{sorensen2018covariate}.

In practice, both methods can be hard to tune due to not having a well defined likelihood.  In the GLM case, CLasso and GMUS both use an elbow-plot, as explained in \cite{sorensen2018covariate}, to determine the amount of regularization, which requires user input and produces unclear results.  While GMUS is a convex optimization, the CLasso is not.  Originally, \cite{loh2011high} show favorable convergence properties in the Gaussian residual case, we find that the GLM solution from CLasso in a popular implementation does not always share these nice properties.  Finally, we note that CLasso and the proposed method require some knowledge of the measurement error variability, whereas GMUS does not.  However, in many applications, like gene expressions, replicates are taken with common practice, which allows for estimation of the variability of the contamination.  Hence, lacking the ability to incorporate the measurement error variability could be a disadvantage for various settings.

\subsection{Overview}

The outline for the remainder of this work is as follows.  In Section \ref{IRO-GLM-ME}, we establish the additive measurement error formulation and the IRO-algorithm.  We show how the IRO-algorithm can be used in solutions pertaining to the context of contaminated linear models and we give practical considerations for its usage.  Section \ref{IRO-GLM-Compute} establishes required imputation procedures for continuous, categorical, and count data.  This is done by assuming the response has parametric form of Gaussian, binomial, and negative binomial distributions, respectively.  A simulation study is then presented in Section \ref{IRO-GLM-Simulation}, illustrating our method's performance in Gaussian and binomial linear regression.  Finally, a data analysis is presented in Section \ref{IRO-GLM-Data}, illustrating the proposed method with two other correction procedures on an experiment using microarray gene expressions to find underlying causes of a tumor relapsing.  All derivations and further results can be found in the Appendix.

\section{The IRO-Algorithm for EIV Regression} \label{IRO-GLM-ME}
Consider the following additive measurement error formulation that will persist for the remainder of the paper.  Let $\bmX = (\bmx_1 , \ldots , \bmx_n)^T \in \mathbb{R}^{n \times p}$ be $n$ independent and identical realizations of a $p$-dimensional random variable, where, for covariance matrix $\bm\Sigma_x$, $\bmx_i \sim N(\bm0_p , \bm\Sigma_x)$.  Instead of directly observing realization $\bmx_i$, we observe $r_i \geq 2$ contaminated replicates.  Assume the contaminated observation to be related additively to the true realization, where 
\begin{equation}
\bmw_{ij} = \bmx_i + \bmu_{ij}
\label{eq:contam}
\end{equation}
such that $\bmu_{ij} \sim N(\bm0_p , \bm\Sigma_u)$ for $i = 1 , \ldots , n$ and $j = 1 , \ldots , r_i$; note that, by independence of $\bmx_i$ and $\bmu_{ij}$, that $\bmw_{ij} \sim N(\bm0_p , \bm\Sigma_x + \bm\Sigma_u)$.  Denote the collection of observation $i$'s replicates as $\bmW_i = (\bmw_1 , \ldots , \bmw_{r_i})^{T}$, and let $\bar{\bmw}_i$ to be the average of the replicates for observation $i$.  Without loss of generality, we assume the measurement error is centered at $\bm0$.

\subsection{The Imputation Regularization Optimization Algorithm} \label{IRO-outline}

The Imputation-Regularized Optimization (IRO) algorithm was recently introduced in the context of high-dimensional variable selection with missing data \cite{liang2018imputation}.  The IRO-algorithm provides a much needed procedure for imputation in case where $n < p$, as common methods, like the well known EM-algorithm \cite{dempster1977maximum}, can fail due to inconsistent or non-unique likelihoods \cite{yi2015regularized}. The IRO-algorithm consists of two iterative steps. At iteration $t = 1 , \ldots , T$, missing values, $\bm{z}_m$, are imputed through a predictive density that is conditioned on the observed data, $\bm{z}_o$, and the estimated model parameters from the previous iteration, $\bm\Delta^{(t-1)}$, namely 
\begin{equation}
\bm{z}_m^{(t)} \sim \pi\left(\bm{z}_m \vert \bm{z}_o , \bm\Delta^{(t-1)}\right).
\end{equation}
The newly generated values $\bm{z}_m^{(t)}$ are then used with observed values $\bm{z}_o$ to estimate the model parameters with a regularized objective function
\begin{equation}
\bm\Delta^{(t)} = \argmin_{\bm\Delta} \left\{ F(\bm\Delta ; \bm{z}_m^{(t)} , \bm{z}_o) + P(\bm\Delta ; \lambda) \right\}, 
\label{eq:o-step}
\end{equation}
where $F$ denotes the parameter's relationship to the data and $P$ denotes the regularization function with sparsity parameter $\lambda$.  The two steps are iterated, and, when the optimization in \eqref{eq:o-step} is asymptotically consistent, the IRO-algorithm forms a valid Markov chain that provides an asymptotically consistent estimate of the high-dimensional variables under mild conditions \cite{liang2018imputation}.

\subsection{The I-Step for High-Dimensional EIV Regression} \label{I-Step-ME}

Measurement error is similar in nature to missing data in that the missing values are related by some underlying density like the contaminated variable.  Regardless of procedure, the conditional density for $\bmx_i \vert \bmW_i , \bm\Omega_u , \bm\Omega_x$ must be estimated for all $i = 1 , \ldots , n$.  Recently, the IRO-algorithm was used in a measurement error correction procedure for Gaussian graphical models, which estimates the precision matrix $\bm\Omega_x$ with an assumed known or estimable $\bm\Omega_u$ \cite{Byrd2019}.  Going forward we will refer to a procedure using the IRO-algorithm to correct for measurement error as an IRO-adjusted procedure. Referring back to the aforementioned contaminated model in \eqref{eq:contam}, the predictive density used to compute the imputation is the full conditional found in Normal-Normal models in Bayesian inference,
\begin{equation}
\pi(\bmx_i \vert \bmW_i , \bm\Omega_u , \bm\Omega_x) \sim N(r_i \bm\Lambda\bm\Omega_u\bar{\bmw}_i , \bm\Lambda),
\label{eq:imputation-onlyx}
\end{equation}
where $\bm\Lambda = (\bm\Omega_x + \bm\Omega_u)^{-1}$, as shown in the Appendix \ref{appendix-derivations-ggm}.

We consider the high-dimensional EIV regression problem for GLMs with response $y \sim \mathcal{D}$ and nuissance parameters $\bm\Theta$.  Here, we assume a relationship exists between the expectation of the response, $y_i$, and a function of the linear combination of covariates, $\bmx_i$. Namely, we formulate the model
\begin{equation}
\mathbb{E}_{\mathcal{D}}(y_i ; \bm\Theta) = f(\bmx_i^{T} \bm\beta)
\end{equation}
for the inverse-link function $f$ and sparse coefficients $\bm\beta \in \mathbb{R}^p$.  The sparsity of the coefficients implies that most are 0.  Denote the number of non-zero coefficients by $q = \vert \vert \bm\beta \vert \vert_0$, where $q \leq n$ and, typically, $q \ll p$.  Instead of observing pair $(y_i , \bmx_i)$, the covariate is observed with (replicated) contamination $(y_i , \bmW_i)$.  To implement the IRO-algorithm for the EIV regression problem the imputation step in \eqref{eq:imputation-onlyx} must be adjusted to include the response model.  The imputation distribution is altered, up to a normalizing constant, as 
\begin{equation}
\pi(\bmx_i \vert y_i , \bmW_i , \bm\Omega_x ,  \bm\Omega_u , \bm\beta , \bm\Theta) \propto 
\pi(y_i \vert \bmx_i , \bm\beta , \bm\Theta) \pi(\bmx_i \vert \bmW_i , \bm\Omega_x , \bm\Omega_u),
\label{eq:imputation-glm}
\end{equation}
where the distribution of $\bmx_i \vert \bmW_i , \bm\Omega_u , \bm\Omega_x$ is as in \eqref{eq:imputation-onlyx}.  For well specified densities for each function, the distribution for the imputation step in \eqref{eq:imputation-glm} is known and easily sampled, which will be explored in Section \ref{IRO-GLM-Compute}.  Once $\bmX$ has been imputed, an estimate of $\bm\Omega_x$ and $\bm\beta$ is obtained, and the process repeated.  The general procedure is presented in Algorithm 1.

\begin{algorithm}[h]
\caption{The IRO-adjusted Procedure for Contaminated GLMs} 
\label{alg:IRO-BAGUS}

\begin{algorithmic}[1]
\State Set number of IRO iterations, $T$
\State Input known $\bm\Omega_u$ or obtain $\hat{\bm\Omega}_u$ using replicate data 
\State Obtain initial estimate for $\bm\beta^{(0)}$ and $\bm\Omega_x^{(0)}$ using $\bar{\bmW}$

\For{t = 1, \ldots , T}

\For{i = 1 , \ldots , n} 
\State draw $\bmx_i^{(t)} \sim \pi(\bmx_i \vert y_i , \bmW_i , \bm\Omega_x^{(t-1)} ,  \hat{\bm\Omega}_u , \bm\beta^{(t-1)} , \bm\Theta^{(t-1)})$ \Comment{Impute}
\EndFor

\State Estimate $\bm\Omega_x^{(t)}$ with $\bmX^{(t)}$ 
\State Estimate $\bm\beta^{(t)}$ and $\bm\Theta^{(t)}$ \Comment{Regularize}

\EndFor

\end{algorithmic}
\label{algo:irome}
\end{algorithm}

\subsection{The RO-Step for High-Dimensional EIV Regression} \label{RO-Step-GLM}

Once the imputation step has been performed, then the remaining parameters must be estimated from the imputed realizations.  Beginning with $\bm\Omega_x$, the precision matrix of the true underlying data, it is tempting to estimate the covariance directly and then invert the estimated covariance.  However in the setting where $n < p$, the estimated covariance is likely to not be of full-rank, and hence inversion would not be possible due to $\hat{\bm\Sigma}_x$ being singular.  Additionally, even if one could reasonably estimate $\bm\Sigma_x$, inversion is computationally expensive.  The Gaussian graphical model literature has given several ways to estimate $\bm\Omega_x$ directly with a regularization term to impose sparsity, which could then estimate a full rank matrix \cite{friedman2008sparse}.  

While estimating the off-diagonal elements of $\bm\Omega_x$ is appealing, many regularization procedures assume independence among covariates.  Even if the assumption is not strictly made, few regularization procedures make use of the dependence structure among the covariates; though some exceptions do exist \cite{yu2016sparse}.  Disregarding the dependency between covariates allows for estimation of only the diagonal of $\bm\Omega_x$.  This results in computational gains in the imputation step, as explained in Section \ref{IRO-GLM-Compute}, and saves a costly optimization of $\bm\Omega_x$.  We observe in our simulations with dependent covariates that estimating only the diagonal of $\bm\Omega_x$ performs well.

Many procedures have been developed to estimate coefficients in regularized general linear models \cite{park2007l1}.  Any method which is consistent will be adequate for the regularization step in estimating the coefficients at each iteration.  The more accurate the regularization method, the better the imputation.  Of particular note is the ability to estimate the nuisance parameter $\bm\Theta$, which is required for the imputation step.  This is a known problem in, for instance, Gaussian linear regression, where the underlying model variability affects the selection quality \cite{belloni2011square}.  The general IRO-adjusted procedure for mismeasured random variables is to alternate between imputation, as in equation \eqref{eq:imputation-glm}, and optimizing parameters $\bm\Omega_x , \bm\beta , \text{and } \bm\Theta$.

\subsection{Computational Considerations}

Most of the regularized optimization procedures for GLMs require some hyperparameter tuning.  For example, consider the Lasso penalty's Lagrangian form, then the optimization in \eqref{eq:glm} will be such that $P(\bm\beta ; \lambda) = \lambda \vert \vert \bm\beta \vert \vert_1$.  The hyperparameter $\lambda$ directly affects the output by controlling the amount of sparsity, and hence needs to be tuned.  We handle this hyperparameter tuning at each iteration of the IRO-algorithm via conventional procedures like k-fold cross-validation, which for many competing methods is unavailable.  Using the standard tool set makes analysis for a practitioner easier as measurement error invalidates traditional methods, and additional methods like \cite{datta2019note} are incorporated into the procedure.  Moreover, competing methods are often in a position of tuning a grid of hyperparameters \cite{belloni2016ell_}.  This adds to their computational costs to find an optimal solution, which may not be plausible depending on the difficultly of the optimization.  Moreover, this adds to the difficultly of use for practitioners, and a higher chance of misapplication or misinterpretation.

We briefly note the similarity of the IRO-algorithm and Gibbs samplers from Bayesian literature \cite{smith1993bayesian}.  Gibbs samplers require obtaining  the distribution of each random variable conditioned on the all other random variables in the model, known as the full conditional distribution.  These distributions are then used to generate values of that random variable, conditioned on the most recently generated value of the other random variables.  This is similar to the IRO-algorithm, which replaces sampling of some variables with an optimization step.  As such, the massive amount of literature that has been developed for Gibbs sampling is applicable to procedures using the IRO-algorithm.  This was illustrated in \cite{liang2018imputation}, where the well-known Gelman-Rubin diagnostics \cite{gelman1992inference} were used to illustrate convergence of the IRO-algorithm.  We note that both samplers can take some time to reach reasonable areas of the posterior distribution.  While a good starting value helps, it is still often beneficial to discard some initial amount of iterates as burn-in.  We make use of this practice in our implementation of IRO-algorithm.

The IRO-algorithm estimates a set of coefficients at each iteration, and the differences in the estimated coefficients may be interpreted as the amount of variation added into the estimation process as a result of the contaminated observations \cite{liang2018imputation}.  With mild conditions on the regularization procedure and variability of the data, the findings of \cite{liang2018imputation} show that the IRO-algorithm gives a consistent estimate of the optimized parameters in each iteration.  Moreover, the results of \cite{Byrd2019} extend this result to the measurement error scenario.  A typical final estimate would make use of all iterations, such as taking the average estimated parameter from each iteration.  However, the average of multiple sparse vectors is not guaranteed to be sparse, which does not give an easy interpretation of the variable selection.  Intuitively, spurious coefficients that appear in the model ought to do so a few number of times; therefore a trimmed mean could be used.  Alternatively, we find using the median of each estimated coefficient as the final estimate to give reliable estimates, as illustrated in Section \ref{IRO-GLM-Simulation}. 

\section{IRO-Adjustments for Some Contaminated GLMs} \label{IRO-GLM-Compute}
In this section we explore imputation steps for three common types of response data: continuous, categorical, and count.  This is done by determining the necessary form of Equation \eqref{eq:imputation-glm} for responses distributed as a Gaussian, binomial, and negative binomial distribution.  These distributions are standard for GLMs, and cover most use cases.  We illustrate that the imputation can be accomplished from known, parameterized distributions, which makes the sampling painless.  Additionally, we address computational considerations of the imputation step.  While we focus here on closed form distributions to be used in the imputation step, there may not always be well-known distributional forms available for every class of model.  Many procedures exist for approximating distributions, such as the Integrated Nested Laplace Approximation \cite{rue2009approximate}.  Samples drawn from the output of these methods could be used to estimate unknown distributional forms given by other models.  All derivations are deferred to the Appendix \ref{appendix-derivations}.

\subsection{Gaussian Linear Regression} \label{IRO-GLM-Gaussian}

The natural starting point is continuous data with Gaussian linear regression.  Here, we assume the response follows the familiar model, $y_i \sim N(\bmx_i^{T}\bm\beta , \sigma^2)$ for all $i = 1 , \ldots , n$.  The imputation step can be shown to be
\begin{equation}
\bmx_i \vert y_i , \bm\beta , \bmW_i , \bm\Omega_u , \bm\Omega_x 
\sim N\left( \bm\Lambda_{G} \left(r_i \bm\Omega_u \bar{\bmw}_i + \frac{y_i}{\sigma^2} \bm\beta \right) , \ \bm\Lambda_{G} \right),
\end{equation}
where $\bm\Lambda_{G} = (\bm\Lambda^{-1} + \sigma^{-2} \bm\beta\bm\beta^{T})^{-1}$ for $\bm\Lambda$ as defined in \eqref{eq:imputation-onlyx}.  We note the impact of quality estimates for $\bm\beta$ and $\sigma^2$, which will be used iteratively for the imputations.  Many regularization procedures do not incorporate the residual variability into the estimation.  If one is confident in the quality of the estimates directly from the regularized model, then the residual variance could be estimated as
\begin{equation}
\hat{\sigma}^2 = \frac{\vert \vert \bm{y} - \bmX^{(t)} \bm\beta^{(t)} \vert \vert_{2}^2}{n - \hat{q}},
\end{equation}
where $\hat{q}$ is the number of estimated, non-zero coefficients.  However, many procedures' performance is known to become worse when the model error variance is not 1 \cite{belloni2011square}.  Hence, if using such a method, like Lasso, it is often beneficial to instead use variants that incorporate the error variance, like the scaled Lasso \cite{sun2012scaled}, into the model during the optimization procedure.

We remark on the computation of the imputation step, which requires inverting the sum of a full rank and rank-1 matrix.  If the features are modeled as independent, implying $\bm\Omega_x$ and $\bm\Omega_u$ are diagonal, then some computational gains can be found by noting that $\bm\beta\bm\beta^{T}$ is a rank-1 matrix.  A typical procedure for generating $p$-dimensional Gaussian data is to generate $p$ independent standard Normal variables, and then to multiply this vector by the Cholesky decomposition of the covariance matrix.  The Cholesky decomposition of a diagonal matrix is simply the square root of the diagonal elements, which can then be updated by $\bm\beta\bm\beta^{T}$ in $\mathcal{O}(p^2)$ time instead of $\mathcal{O}(p^3)$ time if done directly.  There is not an easy way to address the problem of generating the imputation step when $\bm\Omega_x$ or $\bm\Omega_u$ is not diagonal without making assumptions on its form. However, this is a well known problem in Bayesian literature, and recent advances, such as \cite{bhattacharya2016fast}, may prove applicable to our situation in the future.

\subsection{Binomial Linear Regression} \label{IRO-GLM-Binomial}

We now consider the binomial linear regression setting where covariates are contaminated with measurement error.  For each observation $i$, let $y_i \sim Bern(p_i)$ such that
\begin{equation}
p_i = \frac{e^{\bmx_i^{T}\bm\beta}}{1 + e^{\bmx_i^{T}\bm\beta}}.
\label{eq:inverse-logit}
\end{equation}
Incorporating binomial regression into the IRO-algorithm is not as immediate as in Gaussian linear regression due to the logit function, which maps the linear combination of the covariates to the success probability.  The Gaussian setting is conjugate, and hence easily found as in Bayesian inference.  Binomial linear regression is known to not have a closed form full conditional distribution due to the logit function, and has been an long-time area of interest in Bayesian literature \cite{holmes2006bayesian}.  Due to the overlap in the IRO-algorithm and Gibbs sampling methodologies, we are able to utilize some of these findings to incorporate into the imputation step.

Specifically, we will make use of a recent advancement in a line of research using data-augmentation to achieve a well-known distribution for the imputation step.  Using a newly proposed P\'olya-Gamma family of distributions, \cite{polson2013bayesian} have been successful in implementing a procedure that allows for a closed-form binomial regression Gibbs sampler. A random variable $z$ is P\'olya-Gamma distributed with parameters $b \in \mathbb{R}^{+}$ and $c \in \mathbb{R}$ if
\begin{equation}
z = \frac{1}{2\pi^2} \sum_{k = 1}^{\infty} \frac{g_k}{(k - 1/2)^2 + c^2/(4\pi^2))},
\label{eq:polya-gamma}
\end{equation}
where $g_k \sim Ga(b,1)$ are \textit{iid} Gamma random variables; we denote the the P\'olya-Gamma distribution as $z \sim PG(b,c)$.  The main result in \cite{polson2013bayesian} is that 
\begin{equation}
\frac{(e^\psi)^a}{(1 + e^\psi)^b} = 2^{-b} e^{\kappa \psi} \int_{0}^{\infty} e^{-z\psi^2/2} \pi(z) dz,
\label{eq:polya-gamma}
\end{equation}
where $\kappa = a - b/2$ and $z \sim PG(b , 0)$.  Note that when $\psi = \bmx_i^{T} \bm\beta$, the integrand of \eqref{eq:polya-gamma} is the kernel of a Gaussian distribution with respect to $\bmx_i^{T} \bm\beta$. Hence, the inverse-logit function, as in \eqref{eq:inverse-logit}, can be expressed as an infinite convolutional mixture of normal and gamma distributions.

Exploiting the mixture representation of the logit function in \eqref{eq:polya-gamma}, \cite{polson2013bayesian} showed that a Gibbs sampler was possible by exploiting the Normal-Normal conjugacy of the prior on the coefficients and Gaussian kernel.  This procedure is possible by including the P\'olya-Gamma random variable into the sampler.  Thus, in addition to needing to impute $\bmX$, our imputation step must also sample $\bm{z} = (z_1 , \ldots , z_n)^{T}$, which requires the full conditional distribution of $z_i$.  Fortunately, sampling $z_i \vert \bmx_i , \bm\beta$ is an easy task.  \cite{polson2013bayesian} showed that 
\begin{equation}
z_i \vert \bmx_i , \bm\beta \sim PG(1 , \bmx_i^{T} \bm\beta),
\end{equation}
which has been illustrated to have an efficient sampling routine \cite{polson2013bayesian}.  The full conditional to sample each observation's true realization is then Gaussian, namely
\begin{equation}
\bmx_i \vert y_i , z_i , \bm\beta , \bmW_i , \bm\Omega_u , \bm\Omega_x 
\sim N\left( \bm\Lambda_{B} \left(\kappa_i \bm\beta + r_i \bm\Omega_u \bar{\bm{w}}_i \right) , \bm\Lambda_{B} \right)
\label{eq:binomial-impute}
\end{equation}
where $\kappa_i = y_i - 1/2$ and $\bm\Lambda_B = (\bm\Lambda^{-1} + z_i \bm\beta \bm\beta^{T})^{-1}$.  This computation is facilitated by assuming $\bmX$ to be normally distributed, as in the Gaussian linear regression case.

Hence the IRO-algorithm in this context will alternate between imputing $\bmX$ and $\bm{z}$, then optimizing regression coefficients $\bm\beta$ and the covariate's precision $\bm\Omega_x$.  While we have focused on the binomial case, the P\'olya-Gamma augmentation can be extended to the multinomial linear regression case \cite{polson2013bayesian} \cite{chen2013scalable} \cite{linderman2015dependent}.  The inclusion of $\bm{z}$ was shown to create an uniformaly ergodic Gibbs sampler \cite{choi2013polya}, and similar logic should apply to the IRO-adjusted procedure.  Additionally, when $\bm\Omega_x$ and $\bm\Omega_u$ are assumed to be diagonal, a similar procedure to the Gaussian linear case can be used to quickly sample from the Normal distribution in \eqref{eq:binomial-impute}.  Unfortunately, this procedure will need to be computed $n$ times, for each coefficient, as the inverse requires observation specific $z_i$.
\subsection{Negative Binomial Linear Regression} \label{IRO-GLM-NegBin}

Finally, we briefly consider response data being observed as counts.  In the GLM framework, the typical procedures for modeling count data are variants of Poisson and negative binomial regression.  We opt for the more flexible of the two methods, negative binomial regression, which is less susceptible to overdispersion by not enforcing the mean and variance to be the same.  Remarkably, the P\'olya-Gamma augmentation works for any distribution in the binomial family, and hence the negative binomial imputations can be implemented in similar nature to Section \ref{IRO-GLM-Binomial}.  The full conditional for the imputing $\bmx_i$ is exactly as in \eqref{eq:binomial-impute}.  However, the augmented variable $\bm{z}$ is of slightly different form.  Appealing to the additive nature of the prior distribution, as in \eqref{eq:polya-gamma}, for $y_i$ observed counts out of $m_i$ trials, then 
\begin{equation}
z_i \vert \bmx_i , \bm\beta , m_i \sim PG(m_i , \bmx_i^{T} \bm\beta)
\end{equation}
as shown in \cite{polson2013bayesian}.  While sampling the full conditional density becomes more costly as $m_i$ grows, efficient routines have been explored to quickly generate samples \cite{polsonimproved}.

\section{Simulation} \label{IRO-GLM-Simulation}
Here, we examine the numerical performance of the our proposed estimator for high-dimensional Gaussian and binomial linear regression under different settings.  In each setting, five different estimates are compared in terms of estimation quality and variable selection.  The first two estimates come from running the same regularization procedure used in the IRO-adjustment, the MCP penalty \cite{zhang2010nearly}, on (1) the true realizations (Ideal) and (2) the average of the contaminated replicates for each realization (Naive).  The MCP was also then used in (3) our implementation of the IRO-algorithm for measurement error (IRO).  In addition to comparing the performance to the ideal and naive model, we also inspect two other competing models: (4) the Corrected Lasso (CLasso) \cite{loh2011high} and (5) the Generalized Matrix Uncertainty Selector (GMUS) \cite{sorensen2018covariate}, as described in Section \ref{IRO-GLM-LitReview}.  

All computations were performed in R.  To illustrate the ease of incorporating established methodologies into the IRO-adjustment, we make use of a standard regularization package.  The MCP penalty was implemented with the R package `ncvreg'.  This package has been developed using efficient coordinate-descent algorithms created for non-convex regularization, and is built with care to appropriately handle possible numerical issues in the optimization.   For model tuning, the MCP procedure used the package default 10-fold cross-validation.  The Corrected Lasso and GMUS procedures were implemented with the R package `hdme'.   For model tuning, the Corrected Lasso is able to take advantage of cross-validation, and used 10-fold cross-validation for tuning.  However, the GMUS procedure requires hand-tuning for each problem by inspecting a scree-plot and choosing the point where the number of zero coefficients stabilizes.  We automate this tuning for the simulation study by choosing the the first tuning parameter such that the following two points in the grid give the same number of non-zero coefficients.

For each setting, one of two different sets of coefficients are inspected.  The two sets of coefficients are as follows: 
\begin{enumerate}
\item $\bm\beta_1^* = (1, \ldots , 1 , -1 , \ldots , -1 , 0 , \ldots , 0)^{T}$ where 1 and -1 are repeated 5 times with all $p - 10$ remaining coefficients set to 0,

\item $\bm\beta_2^* = (1 , 1/2 , 1/3 , \ldots , 1/10 , 0 , \ldots , 0)^{T}$ where, again, $p - 10$ coefficients are set to 0.
\end{enumerate}
The measurement error was generated from a $\bm0$ mean Gaussian distribution, with diagonal covariance $\bm\Sigma_u$. To control for the signal-to-noise ratio, we use $\gamma \in \{0.5 , 1\}$ as $\text{diag}(\bm\Sigma_u) = \gamma \text{diag}(\bm\Sigma_x)$.  Each observation in every case was generated to have $r = 3$ replicates.  Each setting, as described in the following sections, was implemented with $n = 400$, p = $\{100, 500, 1000\}$, and 100 random instances.  Additionally, the IRO-algorithm ran for $T = 100$ imputation steps. To inspect the performance of each model, we take the average of the $\ell_2$-norm difference (L2) of the estimated and true coefficients from each replicate within each setting,
\[
\ell_2(\hat{\bm\beta}) = \frac{1}{100} \sum_{i=1}^{100} \vert \vert \hat{\bm\beta}_i - \bm\beta^* \vert \vert_2^2,
\]
which measures the quality of the estimated coefficients.  The variable selection quality is reported by the average number of true positives (TP) and false positives (FP).  

\subsection{Gaussian Linear Regression} \label{IRO-Gaussian-Simulation}
We begin by examining Gaussian linear regression.  In addition to the MCP penalty, we also inspected the performance using the Scaled Lasso \cite{sun2012scaled}, for which we defer discussion and results to the Appendix \ref{appendix-scalelasso}.  Three different data generating processes were considered, where data is generated such that $\bmX \sim N(\bm0_p , \bm\Sigma_x)$ and $y_i = \bm{x}_i^{T} \bm\beta + \epsilon_i$ for $\epsilon_i \sim N(0 , \sigma^2).$  The three settings inspect different values of $\bm\Sigma_x , \bm\beta, \text{and } \sigma^2$, and are given by the following:
\begin{enumerate}[label=(G\arabic*)]
\item $\bmX \sim N(\bm0 , \bm\Sigma_x)$ such that the covariance is diagonal where $\bm\Sigma_x = \bm{I}$.  We use $\bm\beta_2^*$ to define the relationship $\bm{y} = \bmX \bm\beta_2^* + \bm\epsilon$, where $\bm\epsilon \sim N(\bm0 , \bm{I})$.

\item $\bmX \sim N(\bm0 , \bm\Sigma_x)$ such that the covariance is diagonal where $\bm\Sigma_x = \bm{I}$. We use  $\bm\beta_1^*$ to define the relationship $\bm{y} = \bmX \bm\beta_1^* + \bm\epsilon$, where $\bm\epsilon \sim N(\bm0 , 3 \bm{I})$.

\item $\bmX \sim N(\bm0 , \bm\Sigma_x)$ such that $\bm\Omega_x = \bm\Sigma_x^{-1}$ is generated with a band structure so that the diagonal and super-diagonal elements are non-zero. This is generated using the `huge' package for the default ``band" setting.  The final covariance has $\text{diag}(\bm\Sigma_x) = \bm1_p$ and a decreasing relation for variables that are further away from each other. The off-diagonal elements have a magnitude starting between 0.4 and 0.55 depending on $p$.  We use $\bm\beta_1^*$ to define the relationship $\bm{y} = \bmX \bm\beta_1^* + \bm\epsilon$, where $\bm\epsilon \sim N(\bm0_p , \bm{I})$.
\end{enumerate}
These settings give potiential situations that arise in practice. 

% latex table generated in R 3.4.2 by xtable 1.8-4 package
% Tue Oct 15 23:47:59 2019
\begin{table}[t]
\centering
\begin{tabular}{ lll | ll | lll }
  \hline
Setting & p & Metric & Ideal & Naive & IRO & CLasso & GMUS \\ 
  \hline
  \multirow{6}{*}{G1} &  \multirow{3}{*}{500} & L2 & 0.319 & 0.405 & \textbf{0.373} & 0.61 & 0.654 \\ 
   &  & TP & 7.39 & 6.2 & \textbf{5.68} & 6.43 & 3.83 \\ 
   &  & FP & 9.53 & 8.47 & 3.02 & 15 & \textbf{0.23} \\ \cline{2-8}
   & \multirow{3}{*}{1000}  & L2 & 0.338 & 0.423 & \textbf{0.391} & 0.62 & 0.676 \\ 
   &  & TP & 7.15 & 6.32 & 5.5 & \textbf{6.07} & 3.61 \\ 
   &  & FP & 11.62 & 11.09 & 3.49 & 18.14 & \textbf{0.34} \\ \hline \hline
   \multirow{6}{*}{G2}& \multirow{3}{*}{500} & L2 & 0.363 & 0.68 & \textbf{0.458} & 1.227 & 1.89 \\ 
   &  & TP & 10 & 10 & \textbf{10} & \textbf{10} & 9.93 \\ 
   &  & FP & 3.43 & 6.78 & 1.02 & 19.34 & \textbf{0.21} \\ \cline{2-8}
   &\multirow{3}{*}{1000}  & L2 & 0.359 & 0.679 & \textbf{0.426} & 1.14 & 1.949 \\ 
   &  & TP & 10 & 10 & \textbf{10} & \textbf{10} & 9.93 \\ 
   &  & FP & 5.13 & 10.54 & 0.98 & 22.44 & \textbf{0.2} \\ \hline \hline 
     \multirow{6}{*}{G3}& \multirow{3}{*}{500} & L2 & 0.424 & 1.138 & \textbf{0.916} & 3.24 & 2.793 \\ 
   &  & TP & 10 & 9.9 & \textbf{9.79} & 7.58 & 5.37 \\ 
   &  & FP & 4.64 & 14.87 & 3.83 & 9.87 & \textbf{0.97} \\ \cline{2-8}
   & \multirow{3}{*}{1000} & L2 & 0.445 & 1.229 & \textbf{1.047} & 3.239 & 2.836 \\ 
   &  & TP & 10 & 9.81 & \textbf{9.58} & 7.27 & 4.94 \\ 
   &  & FP & 8.07 & 23.65 & 4.5 & 12.1 & \textbf{1.49} \\ 
   \hline
\end{tabular}
\caption{Simulation results for Gaussian linear regression under the three specified settings with noise-to-signal ratio $\gamma = 0.5$.  Ideal, Naive, and IRO use the MCP penalty for regularization.  Bold numbers illustrate the best method between the correction procedures for the setting metric.}
\label{table:gaussian-mcp.5}
\end{table}

We display the results for $p = 500$ and $p = 1000$ for settings G1, G2, and G3 when using the MCP penalty for $\gamma = 0.5$ in Table \ref{table:gaussian-mcp.5}; results for $p = 100$ are similar and presented in Appendix \ref{appendix-mcp-full}  To begin, we  compare the results of the Ideal and Naive model to the results of the our IRO-adjusted model.  Focusing on variable selection, it is easy to see that the Ideal model outperforms the Naive model in every setting, as expected.  When comparing the Naive and IRO-adjusted procedure, the biggest take-away is the difference in the number of false positives.  In most every setting the Naive model finds about five times as many false positives.  In all but one setting, the Naive model has a precision considerably less than 0.5.  The IRO-adjusted procedure, however, never falls below 0.6.  The corrected procedure does have more trouble identifying true positives, but the number of true positives never decreases by more than 10\%.  Finally, the quality of the estimated coefficients, as estimated by the norm difference, is always favorable to the IRO-adjusted procedure.

Now, comparing the IRO-adjusted model with the Corrected Lasso and GMUS gives more varied results.  The Corrected Lasso seems to generally have a higher false positive rate and lower true positive rate than both IRO and GMUS.  Interestingly, in setting G2, the Corrected Lasso performs worse than the Naive model, suggesting a lack of robustness to model assumptions.  GMUS does not seem to have much issue at all with false positives, having the lowest amount for every setting.  However, the IRO-adjustment always has more true positives identified.  This can be attributed to using the covariance structure information that GMUS does not take into account.  The IRO-adjusted model and Corrected Lasso have comparable true positive identification.  The IRO-adjusted model appears to have the highest quality coefficient estimates, as illustrated by the superior norm difference of the estimated coefficients in every setting.

\subsection{Binomial Linear Regression} \label{IRO-Binomial-Simulation}
We now consider the case of using binomial linear regression to measure the the relationship of contaminated covariates with binary response.  To this end we consider two settings for this scenario, easily described as the covariates being either independent or dependent.  These settings are:
\begin{enumerate}[label=(B\arabic*)]
\item $\bmX \sim N(\bm0 , \bm\Sigma_x)$ such that the covariance is diagonal where $\bm\Sigma_x = \bm{I}$.  We use $\bm\beta_2^*$ to generate the relationship $y_i \sim Binom(f(\bmx_i^{T} \bm\beta))$.
\item $\bmX \sim N(\bm0 , \bm\Sigma_x)$ such that dependencies exist between features, where $\bm\Sigma_x$ is generated as in setting G3.  We use $\bm\beta_1^*$ to generate $y_i \sim Binom(f(\bmx_i^{T} \bm\beta)).$
\end{enumerate}
In both instances $f$ defines the inverse logit function.  

In Table \ref{table:binomial-results} we display the results of these two settings for $p = 100, 500, \text{and } 1000$ when signal-to-noise is specified such that $\gamma = 0.5$.  Again, we begin by comparing the averaged results of the Ideal and Naive model with the IRO-adjusted procedure that is proposed.  Again, it should be of no surprise that the naive implementation generally performs worse than the ideal.  The effect is extreme for this case, but typically the presence of contaminated observations increases the number of false positives and decreases the number of true positives.  The IRO-adjusted procedure is able to achieve nearly the same number of true positives as the naive method, while reducing the number of false positives by more than half in every case.  Strangely, the IRO-adjusted procedure also has fewer false positives than the ideal model for every case.  This can be attributed to the removal of spurious effects when examining the model at each imputation iteration.  Finally, the IRO-adjustment is able to either do as well or better than the naive model in terms of estimate quality, as measured by the norm difference.

Comparing the results of the IRO-adjustment and alternative correction procedures we note similar results as to the Gaussian case.  Beginning with the CLasso, we first note a general poor performance for each of the settings, having a relatively low number of true positives and poor estimation quality.  As the model is less powerful, less identification was done in total, as seen by the low number of false positives, too.  It may be possible to achieve better performance with extensive tuning, but the defaults already search a well specified grid.  A more likely reason for the performance can be attributed to the non-convex optimization that is performed to find the solution.  While the Gaussian case has been specially designed for finding near-optimal solutions, the GLM case in general is much harder, and would make the elbow method used for tuning a challenge if only some tuning parameters found good solutions.

On the other hand, GMUS was able to find reasonable results for each setting.  We see, again, that the IRO-adjustment performs better in identifying the true positives in the model.  This effect is seen best when the covariates are correlated in Setting B2.  However, GMUS does perform better in regards to the number of false positives in the model.  The choice of method would be then given to the practitioner, as both methods perform better than the naive method.  In terms of estimation quality, the IRO-adjustment performs better in every case.  This is consistent with the results from the Gaussian setting, and establishes a general bias from the GMUS procedure.  We note that, unlike CLasso and GMUS, the IRO-adjusted setting is easily established for other classification methods, like Linear Discriminant Analysis, which could be incorporated to improve the variable selection \cite{witten2011penalized}.

% latex table generated in R 3.4.2 by xtable 1.8-4 package
% Sun Oct 27 17:24:00 2019
\begin{table}[ht]
\centering
\begin{tabular}{lll | ll | lll}
  \hline
Setting & p & Metric & Ideal & Naive & IRO & CLasso & GMUS \\ 
  \hline
 \multirow{9}{*}{B1} & \multirow{3}{*}{100}& L2 & 0.754 & 1.204 & \textbf{0.971} & 3.939 & 2.576 \\ 
   &  & TP & 10 & 10 & \textbf{9.98} & 2.71 & 9.53 \\ 
   &  & FP & 4.16 & 5.9 & 2.45 & \textbf{0.1} & 0.51 \\ \cline{2-8}
   &\multirow{3}{*}{500}  & L2 & 0.898 & 1.537 & \textbf{1.207} & 4.012 & 2.739 \\ 
   &  & TP & 10 & 9.97 & \textbf{9.93} & 2.39 & 9.18 \\ 
   &  & FP & 8.75 & 11.35 & 4.31 & \textbf{0.08} & 0.55 \\ \cline{2-8}
   &\multirow{3}{*}{1000} & L2 & 1.035 & 1.682 & \textbf{1.413} & 4.088 & 2.796 \\ 
   &  & TP & 9.99 & 9.93 & \textbf{9.86} & 2.2 & 8.66 \\ 
   &  & FP & 12.27 & 15.59 & 6.16 & \textbf{0.08} & 0.9 \\ \hline \hline
 \multirow{9}{*}{B2} & \multirow{3}{*}{100} & L2 & 0.728 & 1.528 & \textbf{1.268} & 3.853 & 2.89 \\ 
   &  & TP & 9.97 & 9.57 & \textbf{9.45} & 2.13 & 5.86 \\ 
   &  & FP & 5 & 8.31 & 3.55 & \textbf{0.14} & 1.32 \\ \cline{2-8}
   & \multirow{3}{*}{500} & L2 & 0.934 & 2.132 & \textbf{2.185} & 3.716 & 2.982 \\ 
   &  & TP & 9.81 & 7.85 & \textbf{7.33} & 1.68 & 4.62 \\ 
   &  & FP & 13.65 & 15.15 & 6.16 & \textbf{0.11} & 2.52 \\ \cline{2-8}
   &\multirow{3}{*}{1000}  & L2 & 1.206 & 2.455 & \textbf{2.523} & 3.446 & 3.008 \\ 
   &  & TP & 9.41 & 6.77 & \textbf{5.92} & 1.52 & 4.28 \\ 
   &  & FP & 18.82 & 16.96 & 6.71 & \textbf{0.06} & 2.99 \\ 
   \hline
\end{tabular}
\caption{Simulation results for Binomial linear regression under the two specified settings with signal-to-noise ratio $\gamma = 0.5$, as described in-line.  The Ideal, Naive, and IRO procedures use the MCP penalty for regularization.}
\label{table:binomial-results}
\end{table}

\section{Data Analysis} \label{IRO-GLM-Data}
To show the efficacy of our proposed method, we illustrate it with an application to a microarray gene expression data set.  For the sake of comparison, the data set and the preprocessing steps are the same as in \cite{sorensen2015measurement}.  This data set is comprised of $n = 144$ subjects' gene expressions for favorable histology Wilms tumors, of which 53 relapsed and 91 did not; the data set is accessible by the GEO website, dataset GSE10320 \cite{huang2009predicting}.  Each subject was measured with 10 or 11 probes, and hence replicates are available to estimate the measurement error variability.  The Bioconductor package `bgx' is able to incorporate the subject-level replicates in the preprocessing step to obtain the estimated measurement error covariance \cite{hein2005bgx}, which is assumed diagonal.  To cut down the number of genes to be inspected, any gene that had estimated signal-to-noise value $\gamma > 0.5$ was discarded, the rational being that with too much noise, no discernible selection would be possible, regardless of correction.

We make use of the already processed output for the same dataset found in \cite{nghiem2018simulation} and \cite{Byrd2019}.  After removing genes with estimated signal-to-noise ratio larger than $0.5$, there were a remaining $p = 2074$ genes remaining.  The goal is to determine any genes that have an impact on the tumor relapsing.  We accomplish this with a binomial linear regression.  Similar to the analysis done in Section \ref{IRO-GLM-Simulation}, we compare the results of the Naive estimate, the IRO-adjusted estimate, the CLasso estimate, and the GMUS estimate.  For the purposes of illustration, we use the Lasso procedure with 10-fold cross-validation for the Naive and IRO-adjustment.  As the CLasso and GMUS procedures both lack a well-defined likelihood, we utilize the elbow-plot method as in the Binomial Regression simulation in Section \ref{IRO-Binomial-Simulation}.

We present the results of the analysis in Table \ref{table:data-results}, which shows the total number of genes selected for each procedure and the number of overlap between each procedure. Beginning with the results of the Naive analysis, the Lasso procedure selected a total of 35 genes in total.  The IRO-adjusted Lasso procedure selected less than half that of the Naive implementation, finding a total of 14 genes when taking the median of each iteration's estimated coefficients.  Additionally, all 14 variables found by the IRO-adjusted Lasso were also found by the Naive procedure.  This is in line with the results found in the simulation conducted in the previous section, where the Naive and IRO implementations typically had similar amounts of true positives and a disparate amount of false positives.  

Turning to the competing methods, the CLasso selected a total of 3 genes, all of which are shared with the Naive and IRO-adjustment.  Given the relatively low number of true and false positives in the simulation, this seems to indicate similar behavior.  Finally, the GMUS procedure selected a total of 7 genes.  The behavior of the of GMUS was odd in the sense that there was only one gene in common with the CLasso and two genes in common with the Naive and IRO-adjustment.  We believe that this is likely attributed to the dependencies between the genes, which had a relatively large negative impact on GMUS.  The overall outcome seems to suggest the legitimacy of the simulation study, which illustrated the the IRO-adjustment to be a middle ground between true and false positives.

\begin{table}[ht]
\centering
\begin{tabular}{c | cccc}
& Naive & IRO & CLasso & GMUS \\ \hline
Naive & 35 & 14 & 3 & 2 \\
IRO & 14 & 14 & 3 & 2 \\
CLasso & 3 & 3 & 3 & 1 \\
GMUS & 2 & 2 & 1 & 7 \\ \hline
\end{tabular}
\caption{The total number of selected genes that overlapped between the Naive, IRO-adjusted, CLasso, and GMUS procedures.  Note, the diagonal shows the total number of genes found by each procedure.}
\label{table:data-results}
\end{table}

\section{Conclusion}
We have provided a new method of correction for high-dimensional generalized linear models with regularization.  We employed the recent Imputation Regularization Optimization algorithm in a general correction context, and showed explicitly how to correct for the three most common data types: continuous, categorical, and count.  Our proposed methodology improves on a simple naive implementation, which ignores the measurement error, and is competitive with current existing measurement error correction procedures in this context.  The ease of use is the main draw of our proposal, and does not require special reformulation of existing methods.  This is advantageous for practitioners who can use existing, well designed software, as well as providing an easy way to incorporate new state-of-the-art procedures.  

Future work could be to establish imputation procedures for other settings, such as survival analysis or non-parametric regression.  Many of these settings will not have a well-known density for the imputation step, and hence would require a way of estimating that density for sampling purposes.  Such problems are well-known to Bayesian statisticians, and methods such as the Integrated Nested Laplace Approximation \cite{rue2009approximate} could prove useful.  An alternative direction could be towards establishing post-selection inference procedures on the estimated coefficients.  This notion, termed selective inference, has become popular recently for making a valid inference with regularized models \cite{taylor2015statistical}, and could prove insightful for rigorously providing a final set of estimated coefficients.

\pagebreak

\bibliographystyle{apalike}
\bibliography{ref.bib}

\begin{thebibliography}{}

\bibitem[Belloni et~al., 2011]{belloni2011square}
Belloni, A., Chernozhukov, V., and Wang, L. (2011).
\newblock Square-root {L}asso: pivotal recovery of sparse signals via conic
  programming.
\newblock {\em Biometrika}, 98(4):791--806.

\bibitem[Belloni et~al., 2016]{belloni2016ell_}
Belloni, A., Rosenbaum, M., Tsybakov, A.~B., et~al. (2016).
\newblock An $\{\ell_{1},\ell_{2},\ell_{\infty}\} $-regularization approach to
  high-dimensional errors-in-variables models.
\newblock {\em Electronic Journal of Statistics}, 10(2):1729--1750.

\bibitem[Bhattacharya et~al., 2016]{bhattacharya2016fast}
Bhattacharya, A., Chakraborty, A., and Mallick, B.~K. (2016).
\newblock Fast sampling with {G}aussian scale mixture priors in
  high-dimensional regression.
\newblock {\em Biometrika}, pages 985--991.

\bibitem[Breheny and Huang, 2011]{breheny2011coordinate}
Breheny, P. and Huang, J. (2011).
\newblock Coordinate descent algorithms for nonconvex penalized regression,
  with applications to biological feature selection.
\newblock {\em The Annals of Applied Statistics}, 5(1):232.

\bibitem[Byrd et~al., 2019]{Byrd2019}
Byrd, M., Nghiem, L., and McGee, M. (2019).
\newblock Bayesian regularization of {G}aussian graphical models with
  measurement error.
\newblock {\em arXiv preprint arXiv:1907.02241}.

\bibitem[Carroll et~al., 2006]{carroll2006measurement}
Carroll, R.~J., Ruppert, D., Stefanski, L.~A., and Crainiceanu, C.~M. (2006).
\newblock {\em Measurement Error in Nonlinear Models: a Modern Perspective}.
\newblock Chapman and Hall/CRC.

\bibitem[Chen et~al., 2013]{chen2013scalable}
Chen, J., Zhu, J., Wang, Z., Zheng, X., and Zhang, B. (2013).
\newblock Scalable inference for logistic-normal topic models.
\newblock In {\em Advances in Neural Information Processing Systems}, pages
  2445--2453.

\bibitem[Choi et~al., 2013]{choi2013polya}
Choi, H.~M., Hobert, J.~P., et~al. (2013).
\newblock The {P}olya-{G}amma {G}ibbs sampler for {B}ayesian logistic
  regression is uniformly ergodic.
\newblock {\em Electronic Journal of Statistics}, 7:2054--2064.

\bibitem[Datta and Zou, 2019]{datta2019note}
Datta, A. and Zou, H. (2019).
\newblock A note on cross-validation for lasso under measurement errors.
\newblock {\em Technometrics}, (just-accepted):1--13.

\bibitem[Datta et~al., 2017]{datta2017cocolasso}
Datta, A., Zou, H., et~al. (2017).
\newblock Cocolasso for high-dimensional error-in-variables regression.
\newblock {\em The Annals of Statistics}, 45(6):2400--2426.

\bibitem[Dempster et~al., 1977]{dempster1977maximum}
Dempster, A.~P., Laird, N.~M., and Rubin, D.~B. (1977).
\newblock Maximum likelihood from incomplete data via the {EM} algorithm.
\newblock {\em Journal of the Royal Statistical Society: Series B
  (Methodological)}, 39(1):1--22.

\bibitem[Friedman et~al., 2001]{friedman2001elements}
Friedman, J., Hastie, T., and Tibshirani, R. (2001).
\newblock {\em The Elements of Statistical Learning}, volume~1.
\newblock Springer Series in Statistics New York.

\bibitem[Friedman et~al., 2008]{friedman2008sparse}
Friedman, J., Hastie, T., and Tibshirani, R. (2008).
\newblock Sparse inverse covariance estimation with the {G}raphical {L}asso.
\newblock {\em Biostatistics}, 9(3):432--441.

\bibitem[Gelman et~al., 1992]{gelman1992inference}
Gelman, A., Rubin, D.~B., et~al. (1992).
\newblock Inference from iterative simulation using multiple sequences.
\newblock {\em Statistical Science}, 7(4):457--472.

\bibitem[Hastie et~al., 2015]{hastie2015statistical}
Hastie, T., Tibshirani, R., and Wainwright, M. (2015).
\newblock {\em Statistical Learning with Sparsity: the Lasso and
  Generalizations}.
\newblock Chapman and Hall/CRC.

\bibitem[Hein et~al., 2005]{hein2005bgx}
Hein, A.-M.~K., Richardson, S., Causton, H., Ambler, G.~K., and Green, P.~J.
  (2005).
\newblock Bgx: a fully bayesian gene expression index for affymetrix genechip
  data.
\newblock {\em Biostatistics}, 6(3):349--373.

\bibitem[Holmes et~al., 2006]{holmes2006bayesian}
Holmes, C.~C., Held, L., et~al. (2006).
\newblock Bayesian auxiliary variable models for binary and multinomial
  regression.
\newblock {\em Bayesian Analysis}, 1(1):145--168.

\bibitem[Huang et~al., 2009]{huang2009predicting}
Huang, C.-C., Gadd, S., Breslow, N., Cutcliffe, C., Sredni, S.~T., Helenowski,
  I.~B., Dome, J.~S., Grundy, P.~E., Green, D.~M., Fritsch, M.~K., et~al.
  (2009).
\newblock Predicting relapse in favorable histology {W}ilms tumor using gene
  expression analysis: a report from the {Renal Tumor Committee of the
  Children's Oncology Group}.
\newblock {\em Clinical Cancer Research}, 15(5):1770--1778.

\bibitem[Liang et~al., 2018]{liang2018imputation}
Liang, F., Jia, B., Xue, J., Li, Q., and Luo, Y. (2018).
\newblock An imputation--regularized optimization algorithm for high
  dimensional missing data problems and beyond.
\newblock {\em Journal of the Royal Statistical Society: Series B (Statistical
  Methodology)}, 80(5):899--926.

\bibitem[Linderman et~al., 2015]{linderman2015dependent}
Linderman, S., Johnson, M., and Adams, R.~P. (2015).
\newblock Dependent multinomial models made easy: Stick-breaking with the
  {P}{\'o}lya-{G}amma augmentation.
\newblock In {\em Advances in Neural Information Processing Systems}, pages
  3456--3464.

\bibitem[Loh and Wainwright, 2011]{loh2011high}
Loh, P.-L. and Wainwright, M.~J. (2011).
\newblock High-dimensional regression with noisy and missing data: Provable
  guarantees with non-convexity.
\newblock In {\em Advances in Neural Information Processing Systems}, pages
  2726--2734.

\bibitem[McCullagh, 2019]{mccullagh2019generalized}
McCullagh, P. (2019).
\newblock {\em Generalized Linear Models}.
\newblock Routledge.

\bibitem[Nghiem and Potgieter, 2018]{nghiem2018simulation}
Nghiem, L. and Potgieter, C. (2018).
\newblock Simulation-selection-extrapolation: Estimation in high-dimensional
  errors-in-variables models.
\newblock {\em arXiv preprint arXiv:1808.10477}.

\bibitem[Park and Hastie, 2007]{park2007l1}
Park, M.~Y. and Hastie, T. (2007).
\newblock L1-regularization path algorithm for generalized linear models.
\newblock {\em Journal of the Royal Statistical Society: Series B (Statistical
  Methodology)}, 69(4):659--677.

\bibitem[Perez, 2006]{perez2006preparation}
Perez, J. (2006).
\newblock Preparation of rna for microarray analysis.
\newblock Technical report.

\bibitem[Polson et~al., ]{polsonimproved}
Polson, N.~G., Scott, J.~G., and Windle, J.
\newblock Improved {P}{\'o}lya-{G}amma sampling.
\newblock Technical report.

\bibitem[Polson et~al., 2013]{polson2013bayesian}
Polson, N.~G., Scott, J.~G., and Windle, J. (2013).
\newblock {B}ayesian inference for logistic models using {P}{\'o}lya--{G}amma
  latent variables.
\newblock {\em Journal of the American Statistical Association},
  108(504):1339--1349.

\bibitem[Rocke and Durbin, 2001]{rocke2001model}
Rocke, D.~M. and Durbin, B. (2001).
\newblock A model for measurement error for gene expression arrays.
\newblock {\em Journal of Computational Biology}, 8(6):557--569.

\bibitem[Rosenbaum et~al., 2010]{rosenbaum2010sparse}
Rosenbaum, M., Tsybakov, A.~B., et~al. (2010).
\newblock Sparse recovery under matrix uncertainty.
\newblock {\em The Annals of Statistics}, 38(5):2620--2651.

\bibitem[Rue et~al., 2009]{rue2009approximate}
Rue, H., Martino, S., and Chopin, N. (2009).
\newblock Approximate {B}ayesian inference for latent {G}aussian models by
  using integrated nested {L}aplace approximations.
\newblock {\em Journal of the Royal Statistical Society: Series B (Statistical
  Methodology)}, 71(2):319--392.

\bibitem[Smith and Roberts, 1993]{smith1993bayesian}
Smith, A.~F. and Roberts, G.~O. (1993).
\newblock Bayesian computation via the {G}ibbs sampler and related {M}arkov
  chain {M}onte {C}arlo methods.
\newblock {\em Journal of the Royal Statistical Society: Series B
  (Methodological)}, 55(1):3--23.

\bibitem[S{\o}rensen et~al., 2015]{sorensen2015measurement}
S{\o}rensen, {\O}., Frigessi, A., and Thoresen, M. (2015).
\newblock Measurement error in {L}asso: Impact and likelihood bias correction.
\newblock {\em Statistica Sinica}, pages 809--829.

\bibitem[S{\o}rensen et~al., 2018]{sorensen2018covariate}
S{\o}rensen, {\O}., Hellton, K.~H., Frigessi, A., and Thoresen, M. (2018).
\newblock Covariate selection in high-dimensional generalized linear models
  with measurement error.
\newblock {\em Journal of Computational and Graphical Statistics},
  27(4):739--749.

\bibitem[Sun and Zhang, 2012]{sun2012scaled}
Sun, T. and Zhang, C.-H. (2012).
\newblock Scaled sparse linear regression.
\newblock {\em Biometrika}, 99(4):879--898.

\bibitem[Taylor and Tibshirani, 2015]{taylor2015statistical}
Taylor, J. and Tibshirani, R.~J. (2015).
\newblock Statistical learning and selective inference.
\newblock {\em Proceedings of the National Academy of Sciences},
  112(25):7629--7634.

\bibitem[Tibshirani, 1996]{tibshirani1996regression}
Tibshirani, R. (1996).
\newblock Regression shrinkage and selection via the lasso.
\newblock {\em Journal of the Royal Statistical Society: Series B
  (Methodological)}, 58(1):267--288.

\bibitem[Van~de Geer et~al., 2008]{van2008high}
Van~de Geer, S.~A. et~al. (2008).
\newblock High-dimensional generalized linear models and the {L}asso.
\newblock {\em The Annals of Statistics}, 36(2):614--645.

\bibitem[Witten and Tibshirani, 2011]{witten2011penalized}
Witten, D.~M. and Tibshirani, R. (2011).
\newblock Penalized classification using fisher's linear discriminant.
\newblock {\em Journal of the Royal Statistical Society: Series B (Statistical
  Methodology)}, 73(5):753--772.

\bibitem[Wu et~al., 2008]{wu2008coordinate}
Wu, T.~T., Lange, K., et~al. (2008).
\newblock Coordinate descent algorithms for {L}asso penalized regression.
\newblock {\em The Annals of Applied Statistics}, 2(1):224--244.

\bibitem[Yi and Caramanis, 2015]{yi2015regularized}
Yi, X. and Caramanis, C. (2015).
\newblock Regularized em algorithms: A unified framework and statistical
  guarantees.
\newblock In {\em Advances in Neural Information Processing Systems}, pages
  1567--1575.

\bibitem[Yu and Liu, 2016]{yu2016sparse}
Yu, G. and Liu, Y. (2016).
\newblock Sparse regression incorporating graphical structure among predictors.
\newblock {\em Journal of the American Statistical Association},
  111(514):707--720.

\bibitem[Zhang et~al., 2010]{zhang2010nearly}
Zhang, C.-H. et~al. (2010).
\newblock Nearly unbiased variable selection under minimax concave penalty.
\newblock {\em The Annals of Statistics}, 38(2):894--942.

\end{thebibliography}

\pagebreak

\appendix
\section{Appendix}
\subsection{Derivations} \label{appendix-derivations}

In this section we provide the derivations used to obtain the resulting distributions for the respective imputation steps.

\subsubsection{Covariate Only Imputation Distribution Derivation} \label{appendix-derivations-ggm}

To impute missing true data $\bmx_i$, we wish to find the full conditional distribution of $\bmx_i \vert \bar{\bmw}_i , \bm\Omega_x , \bm\Omega_u$ for each $i = 1 , \ldots , n$. Standard calculations find
\[ \begin{split}
\pi(\bmx_i \vert \bmW_i , \bm\Omega_x , \bm\Omega_u) 
& \propto \exp\left\{-\frac{1}{2} \bmx_i^{T} \bm\Omega_x \bmx_i \right\} 
\prod_{j = 1}^{r_i} \exp\left\{-\frac{1}{2}(\bmw_{ij} - \bmx_{i})^{T} \bm\Omega_u (\bmw_{ij} - \bmx_{i}) \right\} \\
& \propto \exp\left\{ -\frac{1}{2} \left[ \bmx_i^{T} \left( r_ i \bm\Omega_u + \bm\Omega_x \right) \bmx_i - 2 r_i \bmx_i^{T} \bm\Omega_u \bar{\bm{w}}_i \right] \right\} \\
& \propto \exp\left\{ -\frac{1}{2} (\bmx_i - r_i \bm\Lambda \bm\Omega_u \bar{\bmw}_i)^{T} \bm\Lambda^{-1} (\bmx_i - r_i \bm\Lambda \bm\Omega_u \bar{\bmw}_i) \right\},
\end{split} \]
where $\bm\Lambda = (\bm\Omega_x + r_i\bm\Omega_u)^{-1}$.  This result is a kernel of a multivariate Gaussian distribution, 
\begin{equation} 
\pi(\bmx_i \vert \bmW_i , \bm\Omega_x , \bm\Omega_u)  
\sim N(r_i \bm\Lambda \bm\Omega_u \bar{\bmw}_i , \bm\Lambda),
\label{eq:GGM_impute}
\end{equation}
We briefly note that so long as $r_i = r_j$, then observations $i$ and $j$ share the same covariance component.  When generating large multivariate Gaussian distributions, most of the computation comes from the matrix inversions.  By grouping observations with the same number of replicates, time can be saved by only needing compute the full-conditional distributions' covariance once.

\subsubsection{Gaussian Linear Regression Imputation Distribution Derivation} \label{appendix-derivations-gaussian}

The distribution to impute missing data from a linear model with a Gaussian link, where $\bm\Lambda$ is as in \eqref{eq:GGM_impute}, is 
\[ \begin{split}
& \quad \ \pi(\bmx_i \vert \bmW_i , y_i , \bm\Omega_x , \bm\Omega_u , \bm\beta , \sigma^2) 
\propto \pi(y_i \vert \bmw_i , \bm\beta , \sigma^2) \pi(\bmx_i \vert \bmW_i , \bm\Omega_x , \bm\Omega_u) \\
& \propto \exp\left\{ -\frac{1}{2 \sigma^2} (y_i - \bmx_i^{T} \bm\beta)^2 \right\} \exp\left\{ -\frac{1}{2} (\bmx_i - r_i \bm\Lambda \bm\Omega_u \bar{\bmw}_i)^{T} \bm\Lambda^{-1}  (\bmx_i - r_i \bm\Lambda \bm\Omega_u \bar{\bmw}_i) \right\} \\
& \propto \exp\left\{ - \frac{1}{2} \left[ \bmx_i^T (r_i \bm\Omega_u +  \bm\Omega_x + \frac{1}{\sigma^2} \bm\beta \bm\beta^T) \bmx_i - 2\bmx_i^T \left( \frac{y_i}{\sigma^2} \bm\beta + r_i \bm\Lambda \bm\Omega_u \bar{\bmw}_i \right) \right] \right\} \\
& = \exp\left\{ -\frac{1}{2} \left(\bmx_i - \bm\Lambda_{G} \left(r_i \bm\Omega_u \bar{\bmw}_i + \frac{y_i}{\sigma^2} \bm\beta\right)\right)^T \bm\Lambda_{G}^{-1} \left(\bmx_i - \bm\Lambda_{G} \left(r_i \bm\Omega_u \bar{\bmw}_i + \frac{y_i}{\sigma^2} \bm\beta\right) \right) \right\},
\end{split} \]
where $\bm\Lambda_{G} = \left( r_i \bm\Omega_u + \bm\Omega_x + \sigma^{-2} \bm\beta \bm\beta^{T} \right)^{-1}$. This result is, again, the kernel of a multivariate Gaussian distribution, 
\begin{equation}
\pi(\bmx_i \vert \bmW_i , \bm\Omega_x , \bm\Omega_u)  
\sim N\left(\bm\Lambda_{G} \left(r_i \bm\Omega_u \bar{\bmw}_i + \frac{y_i}{\sigma^2} \bm\beta\right) , \bm\Lambda_{G}\right).
\end{equation}
Again, by grouping observations with the same number of replicates, time can be saved by computing each matrix inverse for each unique number of replicates.

\subsubsection{Binomial Linear Regression Imputation Distribution Derivation} \label{appendix-derivations-binomial}

To impute $\bmx_i$ when using the logit function we appeal to \cite{polson2013bayesian}, which, as explained in Section \ref{IRO-GLM-Binomial}, uses P\'olya Gamma random variables to augment the data generating process.  For the most recently generated $z_i$, from \cite{polson2013bayesian} we note that 
\begin{equation}
\pi(y_i \vert z_i , \bmx_i , \bm\beta) 
= \frac{\exp\{\bmx_i^{T} \bm\beta\}^{y_i}}{1 + \exp\{\bmx_i^{T}\bm\beta\}}
\propto \exp\left\{-\frac{z_i}{2} \left(\frac{\kappa_i}{z_i} - \bmx_i^{T} \bm\beta \right)^2 \right\},
\end{equation}
where $\kappa_i = y_i - 1/2$.  Hence, with $\bm\Lambda$ as in \eqref{eq:GGM_impute}, we have
\[\begin{split}
& \quad \pi(\bmx_i \vert \bmW_i , \bm\Omega_x , \bm\Omega_u , y_i , \bm\beta , z_i) \\
& \propto \exp\left\{ -\frac{1}{2} (\bmx_i - r_i \bm\Lambda \bm\Omega_u \bar{\bmw}_i)^{T} \bm\Lambda^{-1} (\bmx_i - r_i \bm\Lambda \bm\Omega_u \bar{\bmw}_i) -\frac{z_i}{2} \left(\frac{\kappa_i}{z_i} - \bmx_i^{T} \bm\beta \right)^2 \right\} \\
& \propto \exp\left\{ -\frac{1}{2} \left[ \bmx_i^{T} (z_i \bm\beta \bm\beta^{T}  + r_i \bm\Omega_u + \bm\Omega_x ) \bmx_i - 2 \bmx_i^{T} (\kappa_i \bm\beta + r_i \bm\Omega_u \bar{\bmw}_i) \right] \right\} \\
& \propto \exp\left\{-\frac{1}{2}(\bmx_i - \bm\Lambda_{B}(\kappa_i \beta + r_i \bm\Omega_u \bar{\bmw}_i))^{T} \bm\Lambda_{B} (\bmx_i  - \bm\Lambda_{B}(\kappa_i \beta + r_i \bm\Omega_u \bar{\bmw}_i)) \right\},
\end{split}\]
where $\bm\Lambda_{B} = (z_i \bm\beta \bm\beta^{T} + r_i \bm\Omega_u + \bm\Omega_x)^{-1}$.  As expected from the results of \cite{polson2013bayesian}, we have a Gaussian kernal, where
\begin{equation}
\pi(\bmx_i \vert \bmW_i , \bm\Omega_x , \bm\Omega_u , y_i , \bm\beta , z_i) \sim 
N(\bm\Lambda_{B}(\kappa_i \beta + r_i \bm\Omega_u \bar{\bmw}_i) , \bm\Lambda_B).
\end{equation}
The derivation for the full conditional distribution of $z_i$ is exactly the same as in \cite{polson2013bayesian}.

\subsection{Estimating the Measurement Error Covariance with Replicates}

Here, we address an estimate of the measurement error's precision matrix, $\bm\Omega_u$, which is necessary for the imputation step.  In some instances, it may be realistic to know the amount of variability in the measurement process; for instance, a machine taking measurements where the output falls within some perturbation of the truth.  However, in many contexts the variability of the contamination process will not be known, and hence need to be estimated, typically with replicates.  Estimating $\bm\Sigma_u$ is difficult due to not directly observing the amount of contamination on each observation.  However, if one assumes the amount of contamination is independent between each variable, then a procedure exists to get an empirical estimate of the diagonal of $\bm\Sigma_x$ and, hence, $\bm\Omega_u$.

An estimate of $\bm\Sigma_u$ is also necessary for the imputation.  When data is observed with replicates for each observation, then this covariance matrix is an estimable variable under independence assumptions.  Consider the measurement error distribution as described in Section \ref{IRO-GLM-ME}, where for each observation's replicates, $\bmu_{ij} \sim N(\bm0_p , \bm\Sigma_u)$.  Estimating $\bm\Sigma_u$ is not trivial because $n < p$ and $\bmu_{ij}$ is not directly observed.  Note for replicate $j$ and $k$ of observation $i$ that
\begin{equation}
\bm{d}_{ijk} = \bmw_{ij} - \bmw_{ik} = \bmx_i - \bmu_{ij} - \bmx_i - \bmu_{ik} = \bmu_{ij} - \bmu_{ik}.
\end{equation}
Assuming the amount of contamination is independent for each covariate, then, for covariate $m$, 
\[
Var(d_{ijk}^{(m)}) = Var(u_{ij}^{(m)}) + Var(u_{ik}^{(m)}) = 2[\bm\Sigma_u]_{m,m}.
\]
Hence, if one were willing to assume the same distribution governing the contamination of each observation, the differences from all $i = 1 , \ldots , n$ where $j < k$ could be used and averaged for all $r_i(r_i-1)$ possible differences per observation,
\begin{equation}
[\hat{\bm\Omega}_u]_{m,m} = \frac{1}{\sqrt{2}} \frac{1}{n} \sum_{i = 1}^{n} \frac{1}{r_i(r_i - 1)} \sum_{j < k} d_{ijk}^{(m)}.
\label{eq:ME_precision}
\end{equation}
If heterogenious measurement error is believed to exist between observations, then it can easily be incorporated into the imputation step by using observation specific $\bm\Omega_{u,i}$. Here, the averaged covariance diagonal element would only be between the pair-wise replicates for the obervation.

\subsection{Other Gaussian Simulation Results} \label{appendix-scalelasso}
\subsubsection{Complete MCP Results} \label{appendix-mcp-full}

In Tables \ref{table:full-mcp-0.5} and \ref{table:full-mcp-1} we display the complete results for the results found in Section \ref{IRO-Gaussian-Simulation}.  Table \ref{table:full-mcp-0.5} displays the results for $p = 100$, and all results are displayed for $\gamma = 1$ in Table \ref{table:full-mcp-1}.  All results are similar to the discussion presented in Section \ref{IRO-Gaussian-Simulation}.

% latex table generated in R 3.4.2 by xtable 1.8-4 package
% Tue Oct 15 23:47:59 2019
\begin{table}[H]
\centering
\begin{tabular}{ lll | ll | lll }
  \hline
Setting & p & Metric & Ideal & Naive & IRO & CLasso & GMUS \\ 
  \hline
\multirow{9}{*}{G1} & \multirow{3}{*}{100}  & L2 & 0.266 & 0.349 & 0.319 & 0.611 & 0.616 \\ 
   &  & TP & 8.45 & 7.61 & 7.09 & 7.57 & 4.34 \\ 
   &  & FP & 5.44 & 4.69 & 2.75 & 9.85 & 0.09 \\ \cline{2-8}
   &  \multirow{3}{*}{500} & L2 & 0.319 & 0.405 & 0.373 & 0.61 & 0.654 \\ 
   &  & TP & 7.39 & 6.2 & 5.68 & 6.43 & 3.83 \\ 
   &  & FP & 9.53 & 8.47 & 3.02 & 15 & 0.23 \\ \cline{2-8}
   & \multirow{3}{*}{1000}  & L2 & 0.338 & 0.423 & 0.391 & 0.62 & 0.676 \\ 
   &  & TP & 7.15 & 6.32 & 5.5 & 6.07 & 3.61 \\ 
   &  & FP & 11.62 & 11.09 & 3.49 & 18.14 & 0.34 \\ \hline \hline
  \multirow{9}{*}{G2} &  \multirow{3}{*}{100}& L2 & 0.33 & 0.602 & 0.422 & 1.293 & 1.6 \\ 
   &  & TP & 10 & 10 & 10 & 10 & 10 \\ 
   &  & FP & 1.33 & 3.15 & 0.75 & 11.06 & 0.16 \\ \cline{2-8}
   & \multirow{3}{*}{500} & L2 & 0.363 & 0.68 & 0.458 & 1.227 & 1.89 \\ 
   &  & TP & 10 & 10 & 10 & 10 & 9.93 \\ 
   &  & FP & 3.43 & 6.78 & 1.02 & 19.34 & 0.21 \\ \cline{2-8}
   &\multirow{3}{*}{1000}  & L2 & 0.359 & 0.679 & 0.426 & 1.14 & 1.949 \\ 
   &  & TP & 10 & 10 & 10 & 10 & 9.93 \\ 
   &  & FP & 5.13 & 10.54 & 0.98 & 22.44 & 0.2 \\ \hline \hline 
   \multirow{9}{*}{G3} & \multirow{3}{*}{100} & L2 & 0.407 & 1.016 & 0.768 & 3.792 & 2.665 \\ 
   &  & TP & 10 & 9.97 & 9.96 & 8.07 & 6.42 \\ 
   &  & FP & 2.03 & 5.51 & 2.02 & 4.36 & 0.61 \\ \cline{2-8}
   & \multirow{3}{*}{500} & L2 & 0.424 & 1.138 & 0.916 & 3.24 & 2.793 \\ 
   &  & TP & 10 & 9.9 & 9.79 & 7.58 & 5.37 \\ 
   &  & FP & 4.64 & 14.87 & 3.83 & 9.87 & 0.97 \\ \cline{2-8}
   & \multirow{3}{*}{1000} & L2 & 0.445 & 1.229 & 1.047 & 3.239 & 2.836 \\ 
   &  & TP & 10 & 9.81 & 9.58 & 7.27 & 4.94 \\ 
   &  & FP & 8.07 & 23.65 & 4.5 & 12.1 & 1.49 \\ 
   \hline
\end{tabular}
\caption{Simulation results for Gaussian linear regression under the three specified settings with noise-to-signal ratio $\gamma = 0.5$.  Ideal, Naive, and IRO use the MCP penalty for regularization.}
\label{table:full-mcp-0.5}
\end{table}

% latex table generated in R 3.4.2 by xtable 1.8-4 package
% Wed Oct 16 23:33:29 2019
\begin{table}[H]
\centering
\begin{tabular}{lll | ll | lll}
  \hline
Setting & p & Metric & Ideal & Naive & IRO & CLasso & GMUS \\ 
  \hline
 \multirow{9}{*}{G1}&  \multirow{3}{*}{100}& L2 & 0.265 & 0.464 & 0.385 & 1.576 & 0.709 \\ 
   &  & TP & 8.38 & 6.73 & 6 & 3.1 & 3.94 \\ 
   &  & FP & 5.79 & 4.95 & 2.17 & 1.37 & 0.12 \\ \cline{2-8}
   &  \multirow{3}{*}{500}& L2 & 0.322 & 0.503 & 0.436 & 1.45 & 0.735 \\ 
   &  & TP & 7.64 & 5.77 & 4.41 & 2.72 & 3.62 \\ 
   &  & FP & 9.37 & 7.69 & 1.02 & 1.8 & 0.22 \\ \cline{2-8}
   &\multirow{3}{*}{1000}  & L2 & 0.342 & 0.524 & 0.456 & 1.408 & 0.762 \\ 
   &  & TP & 6.95 & 5.45 & 4.23 & 2.41 & 3.43 \\ 
   &  & FP & 11.22 & 11.06 & 1.09 & 2.38 & 0.4 \\ \hline \hline 
  \multirow{9}{*}{G2} & \multirow{3}{*}{100} & L2 & 0.351 & 0.913 & 0.535 & 3.943 & 1.937 \\ 
   &  & TP & 10 & 10 & 10 & 7.26 & 9.88 \\ 
   &  & FP & 1.82 & 4.47 & 0.35 & 0.77 & 0.24 \\ \cline{2-8}
   & \multirow{3}{*}{500} & L2 & 0.371 & 1.026 & 0.606 & 3.367 & 2.195 \\ 
   &  & TP & 10 & 10 & 9.99 & 7.28 & 9.62 \\ 
   &  & FP & 3.9 & 9.56 & 0.56 & 1.13 & 0.15 \\ \cline{2-8}
   &\multirow{3}{*}{1000}  & L2 & 0.366 & 1.055 & 0.637 & 3.372 & 2.257 \\ 
   &  & TP & 10 & 10 & 9.99 & 7.07 & 9.62 \\ 
   &  & FP & 4.3 & 13.49 & 0.4 & 1.23 & 0.34 \\ \hline \hline 
  \multirow{9}{*}{G3} & \multirow{3}{*}{100} & L2 & 0.39 & 1.527 & 1.166 & 6.744 & 2.775 \\ 
   &  & TP & 10 & 9.85 & 9.65 & 2.83 & 5.87 \\ 
   &  & FP & 1.66 & 8.69 & 1.89 & 0.05 & 0.74 \\ \cline{2-8}
   & \multirow{3}{*}{500} & L2 & 0.416 & 1.845 & 1.74 & 5.891 & 2.853 \\ 
   &  & TP & 10 & 9.06 & 8.18 & 2.88 & 5.24 \\ 
   &  & FP & 4.87 & 19.58 & 3.12 & 0.14 & 1.76 \\ \cline{2-8}
   & \multirow{3}{*}{1000} & L2 & 0.42 & 2.1 & 2.105 & 5.461 & 2.885 \\ 
   &  & TP & 10 & 8.18 & 6.98 & 3.02 & 4.82 \\ 
   &  & FP & 6.92 & 22.68 & 3.27 & 0.2 & 1.94 \\ 
   \hline
\end{tabular}
\caption{Simulation results for Gaussian linear regression under the three specified settings with noise-to-signal ratio $\gamma = 1$.  Ideal, Naive, and IRO use the MCP penalty for regularization.}
\label{table:full-mcp-1}
\end{table}

\subsubsection{Scaled Lasso Results}

To illustrate the IRO-algorithm with another methodology, we opted to illustrate the incorporation of the Scaled Lasso penalty \cite{sun2012scaled}.  The Scaled Lasso penalty incorporates the residual error term into the Lasso estimation procedure, which is necessary for the imputation step. We display the results in Tables \ref{table-scalelasso.5} and \ref{table-scalelasso1}.  The overall results are similar to the MCP penalty, with a few differences.  One difference is that the Naive model had a difficult time finding convergence.  This occurred in both results for $\gamma = 0.5$ and $\gamma = 1$.  We believe this can be attributed to the residual variability being confused with the covariate mismeasurement variability. This would lead to poor estimation of $\sigma_\epsilon^2$, and hence $\bm\beta$.  The Second differnce is slight degredation of performance for the IRO-correct.  This is likely due to the bias incorporated into the estimate from the $\ell_1$ penalty.  

\begin{table}[ht]
\centering
\begin{tabular}{lll | ll | lll}
  \hline
Setting & p & Metric & Ideal & Naive & IRO & CLasso & GMUS \\ 
  \hline
   \multirow{6}{*}{G1}&  \multirow{3}{*}{500}  & L2 & 0.422 & NA & 0.507 & 0.61 & 0.654 \\ 
   &  & TP & 6.87 & 6.73 & 5.45 & 6.43 & 3.83 \\ 
   &  & FP & 4.79 & 53.81 & 2.23 & 15 & 0.23 \\  \cline{2-8}
   & \multirow{3}{*}{1000} & L2 & 0.448 & NA & 0.542 & 0.62 & 0.676 \\ 
   &  & TP & 6.55 & 6.18 & 5.12 & 6.07 & 3.61 \\ 
   &  & FP & 4.99 & 13.18 & 2.19 & 18.14 & 0.34 \\ \hline \hline
   \multirow{6}{*}{G2}&  \multirow{3}{*}{500}  & L2 & 1.009 & NA & 1.425 & 1.227 & 1.89 \\ 
   &  & TP & 10 & 9.99 & 10 & 10 & 9.93 \\ 
   &  & FP & 5.25 & 107.28 & 2.34 & 19.34 & 0.21 \\  \cline{2-8}
   & \multirow{3}{*}{1000}  & L2 & 1.069 & NA & 1.516 & 1.14 & 1.949 \\ 
   &  & TP & 10 & 10 & 10 & 10 & 9.93 \\ 
   &  & FP & 4.87 & 13.3 & 2.17 & 22.44 & 0.2 \\ \hline \hline
   \multirow{6}{*}{G3}&  \multirow{3}{*}{500}  & L2 & 2.315 & NA & 2.65 & 3.24 & 2.793 \\ 
   &  & TP & 8.99 & 8.28 & 6.72 & 7.58 & 5.37 \\ 
   &  & FP & 4.55 & 42.47 & 2.98 & 9.87 & 0.97 \\ \cline{2-8}
   &  \multirow{3}{*}{1000} & L2 & 2.548 & NA & 2.767 & 3.239 & 2.836 \\ 
   &  & TP & 7.87 & 7.17 & 5.52 & 7.27 & 4.94 \\ 
   &  & FP & 5.15 & 24.31 & 2.62 & 12.1 & 1.49 \\ 
   \hline
\end{tabular}
\caption{Simulation results for Gaussian linear regression under the three specified settings with noise-to-signal ratio $\gamma = 0.5$.  Ideal, Naive, and IRO use the Scaled Lasso penalty for regularization. Due to convergence issues, many L2 norms for the naive method are missing and denoted NA.}
\label{table-scalelasso.5}
\end{table}

\begin{table}[H]
\centering
\begin{tabular}{lll | ll | lll}
  \hline
Setting & p & Metric & Ideal & Naive & IRO & CLasso & GMUS \\ 
  \hline
  \multirow{6}{*}{G1}&  \multirow{3}{*}{500}  & L2 & 0.419 & NA & 0.574 & 1.45 & 0.735 \\ 
   &  & TP & 7.07 & 6.67 & 4.42 & 2.72 & 3.62 \\ 
   &  & FP & 5.01 & 46.3 & 1.42 & 1.8 & 0.22 \\ \cline{2-8}
   & \multirow{3}{*}{1000}  & L2 & 0.453 & NA & 0.614 & 1.408 & 0.762 \\ 
   &  & TP & 6.34 & 5.72 & 4.06 & 2.41 & 3.43 \\ 
   &  & FP & 4.79 & 17.88 & 1.22 & 2.38 & 0.4 \\ \hline \hline
   \multirow{6}{*}{G1}&  \multirow{3}{*}{500}  & L2 & 0.983 & NA & 1.77 & 3.367 & 2.195 \\ 
   &  & TP & 10 & 9.95 & 9.96 & 7.28 & 9.62 \\ 
   &  & FP & 4.92 & 91.09 & 1.51 & 1.13 & 0.15 \\ \cline{2-8}
   &  \multirow{3}{*}{1000} & L2 & 1.057 & NA & 1.938 & 3.372 & 2.257 \\ 
   &  & TP & 10 & 10 & 9.91 & 7.07 & 9.62 \\ 
   &  & FP & 4.8 & 33 & 1.14 & 1.23 & 0.34 \\ \hline \hline
   \multirow{6}{*}{G1}&  \multirow{3}{*}{500}  & L2 & 2.299 & NA & 2.78 & 5.891 & 2.853 \\ 
   &  & TP & 9.06 & 7.98 & 5.44 & 2.88 & 5.24 \\ 
   &  & FP & 5.03 & 65.15 & 1.76 & 0.14 & 1.76 \\ \cline{2-8}
   & \multirow{3}{*}{1000}  & L2 & 2.533 & NA & 2.839 & 5.461 & 2.885 \\ 
   &  & TP & 7.81 & 6.79 & 4.63 & 3.02 & 4.82 \\ 
   &  & FP & 4.63 & 17.58 & 1.58 & 0.2 & 1.94 \\ 
   \hline
\end{tabular}
\caption{Simulation results for Gaussian linear regression under the three specified settings with noise-to-signal ratio $\gamma = 1$.  Ideal, Naive, and IRO use the Scaled Lasso penalty for regularization. Due to convergence issues, many L2 norms for the naive method are missing and denoted NA.}
\label{table-scalelasso1}
\end{table}

\subsection{Other Binomial Regression Results}

We display the results when $\gamma = 1$ for the binomial linear regression simulation found in Section \ref{IRO-Binomial-Simulation}.  Besides Setting B2 being slightly more in favor of the IRO-adjusted procedure, the results are similar.

\begin{table}[ht]
\centering
\begin{tabular}{lll | ll | lll}
  \hline
Setting & p & Metric & Ideal & Naive & IRO & CLasso & GMUS \\ 
  \hline
\multirow{9}{*}{B1} & \multirow{3}{*}{100} & L2 & 0.695 & 1.748 & 1.21 & 3.613 & 2.713 \\ 
   &  & TP & 10 & 9.98 & 9.91 & 2.3 & 9.22 \\ 
   &  & FP & 4.07 & 5.56 & 1.73 & 0.05 & 0.53 \\ \cline{2-8}
   & \multirow{3}{*}{500} & L2 & 0.933 & 1.987 & 1.658 & 3.702 & 2.846 \\ 
   &  & TP & 10 & 9.86 & 9.62 & 2.03 & 8.53 \\ 
   &  & FP & 8.51 & 13.07 & 3.07 & 0.07 & 0.89 \\ \cline{2-8}
   &\multirow{3}{*}{1000}  & L2 & 1.026 & 2.117 & 1.938 & 3.735 & 2.887 \\ 
   &  & TP & 10 & 9.83 & 9.38 & 2.07 & 8.17 \\ 
   &  & FP & 11.86 & 17.51 & 4.08 & 0.12 & 1.41 \\ \hline \hline
   \multirow{9}{*}{B2} & \multirow{3}{*}{100}  & L2 & 0.731 & 2.171 & 2.059 & 3.374 & 2.949 \\ 
   &  & TP & 9.97 & 8.82 & 7.88 & 1.95 & 5.41 \\ 
   &  & FP & 4.99 & 9.23 & 3.11 & 0.1 & 1.6 \\ \cline{2-8}
   & \multirow{3}{*}{500} & L2 & 0.967 & 2.701 & 2.726 & 3.266 & 3.009 \\ 
   &  & TP & 9.74 & 6.18 & 5 & 1.43 & 4.37 \\ 
   &  & FP & 12.8 & 10.83 & 3.35 & 0.05 & 2.38 \\ \cline{2-8}
   &\multirow{3}{*}{1000}  & L2 & 1.216 & 2.825 & 2.859 & 3.276 & 3.026 \\ 
   &  & TP & 9.38 & 5.29 & 4.1 & 1.36 & 4 \\ 
   &  & FP & 17.98 & 12.35 & 3.35 & 0.06 & 3.01 \\ 
   \hline
\end{tabular}
\caption{Simulation results for Binomial linear regression under the two specified settings with signal-to-noise ratio $\gamma = 1$.  The Ideal, Naive, and IRO procedures use the MCP penalty for regularization.}
\end{table}

\subsection{Details for Data Analysis}
\begin{figure}[h]
\centering
\begin{tabular}{c c}
\includegraphics[scale=.35 , trim={1.2cm 0 0 0}]{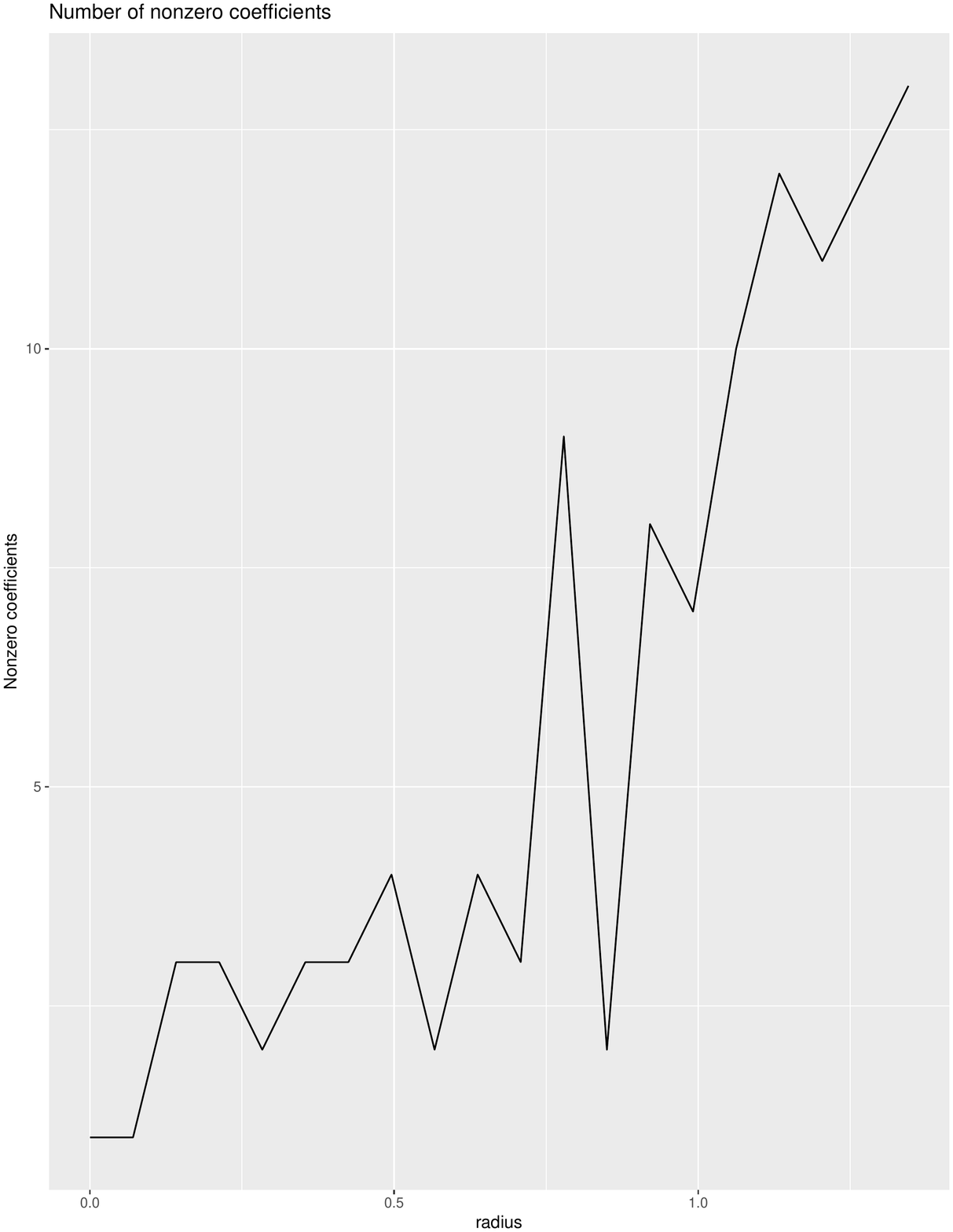} & 
\includegraphics[scale=.35]{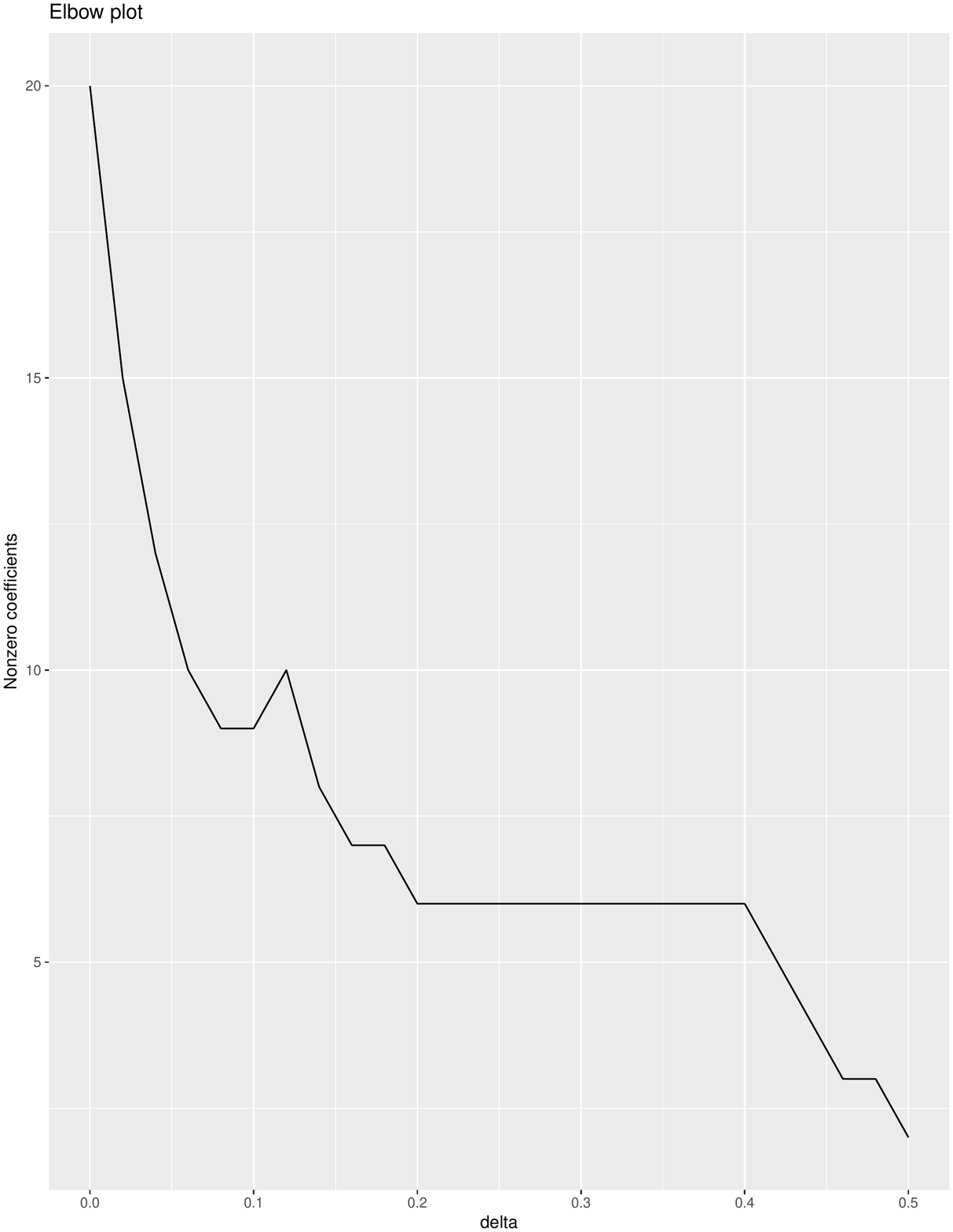}
\end{tabular}
\caption{Outputted ELBO-plots for CLasso (left) and GMUS (right).  Note that the increase of the regularization parameter has varying affect, hence the opposing trend.}
\label{figure:elbo-plot}
\end{figure}

Here we illustrate the ELBO-plots for both the CLasso and GMUS as performed in the data analysis found in Section \ref{IRO-GLM-Data}.  These plots, generated by the `hdme' package output, show the tuning parameter on the $x$-axis and the number of non-zero coefficients on the $y$-axis.  The authors encourage picking where the number of non-zero coefficients stabilize.  That is to say, pick the tuning parameter where the following tuning parameters give the same number of non-zero coefficients.  We present the plots in Figure \ref{figure:elbo-plot}, where the left and right plot is for CLasso and GMUS, respectively.  As noted in Section \ref{IRO-Binomial-Simulation}, the optimality of the solution for CLasso only holds for the Gaussian case, hence the bumpiness. Hence, for CLasso we opt to choose the largest radius value in the grid that gives the number of non-zero coefficients to be 3 as this is the most common amount in a short succession.  GMUS begins to stablize at 0.2, and this is the value used for the analysis.

\end{document}